# Moving an Atom towards Right or Left Side by Applying Quantum Mechanical Matter Wave Near a Surface

Sadia Humaira Salsabil[1], Golam Dastegir Al-Quaderi[2], M.R.C. Mahdy[1*]

[1] *Department of Electrical & Computer Engineering, North South University, Bashundhara, Dhaka 1229, Bangladesh*
[2] *Department of Physics, University of Dhaka, Dhaka 1000, Bangladesh*

[*]Corresponding author's email address: mahdy.chowdhury@northsouth.edu

**The area of trapping the atoms or molecules using light has advanced tremendously in the last few decades. In contrast, the idea of controlling (not only trapping) the movement of atomic-sized particles using quantum mechanical matter waves is a completely new emerging area of particle manipulation. Though a single previous report has suggested the pulling of atoms based on matter wave tractor beams, an attempt is yet to be made to produce a lateral force (moving the atoms towards left or right) using this technique. This article demonstrates a matter wave-based manipulation scenario that gives rise to reversible lateral force on an atom due to the interaction energy of the quantum mechanical matter wave in the presence of a metal surface creating an asymmetrical set-up. For a symmetric set-up, no lateral force has been observed. We have performed several full wave simulations and analytical calculations on a particular set-up of Xenon scatterer atoms placed near a Copper surface, with two plane matter waves of Helium impinging in the direction parallel to the surface from two sides of the scatterer. By solving the time-independent Schrödinger equation and using the solution, quantum mechanical stress tensor formalism has been applied to compute the force acting on the particle. The full wave simulation results have been found in excellent agreement with the analytical calculations. The results for the adsorbed scatterer case suggests that our proposed technique can be an efficient cleaning procedure similar to electron-stimulated desorption for futuristic applications.**

**Introduction**

Optical trapping of micrometer-sized objects has branched out to several methods in the past decades. Laser trapping was popularized by Ashkin in 1970 using two counter-propagating beams [1]. Later in 1986, Ashkin et al. made a revolutionary discovery of trapping using a single beam (optical tweezers) [2]. He also showed the techniques of trapping viruses and living bacteria without harming those [3]. For his contributions, Ashkin was awarded the Nobel Prize in physics in 2018. It is worth mentioning that the laser cooling and trapping have also experienced tremendous advancement in the last few decades. For the development of methods of cooling and trapping atoms with laser light, the 1997 Nobel Prize in Physics was awarded to Claude Cohen-Tannoudji, Steven Chu, and William Daniel Phillips [4].

Our objective in this work is different from trapping of atoms as we are focusing on the controlled motion (i.e., manipulation) of quantum objects such as atoms or molecules, etc. in a lateral direction. In addition, instead of applying light beams, this work focuses on manipulation or control of the atoms (i.e., reversible lateral force) by utilizing matter wave beams [cf. Fig. 1 (a)-(d)]. To the best of our knowledge, so far only one article demonstrated matter wave-based tractor beam, in which the phenomenon of pulling (instead of lateral force) an atomic-sized scatterer by a Bessel beam matter wave was utilized [5].

It is well known that light exerts radiation pressure on an object and it pushes a particle towards in the direction of its propagation by imparting on the particle momentum carried by the quanta of light, photon. In recent years, macroscopic manipulation via light beams has achieved to completely new level of advancement by using various novel techniques. For example, counter intuitive pulling of dielectric or magneto-dielectric objects using light have been demonstrated in the last decade [6-10]. In addition, pulling of atoms using electromagnetic waves has also been reported [11,12]. Yet, very few examples include the optical lateral force [13-17], which acts in the direction normal to the direction of incidence of the light wave.

Some particular properties of the incident electromagnetic wave on the scatterer are responsible for the existence of lateral force. The properties of the incident electromagnetic wave can be attributed to its state of polarization or spin-projection in the case of circularly polarized light [13,14]. Other cases of optical lateral force are mostly connected with the chirality of the material [15-17]. However, these attributes are not directly connected with any quantum mechanical set-up or with our work, since in our work neither the incident particle nor the scatterer possesses any anisotropic property.

Similar to light, quantum mechanical matter wave exerts pressure on an object and it pushes a particle towards the direction of its propagation. This work demonstrates a matter wave-based novel manipulation scenario [cf. Fig. 1 (a)-(d)] that gives rise to reversible lateral force on an atom, due to interaction of the quantum mechanical matter wave with a metal surface, which creates the required asymmetry. For a symmetric set-up, no lateral force has been observed. We have performed several full wave simulations on a particular set-up of Xenon scatterer atoms placed near the Copper surface, with two plane matter waves of Helium impinging in the direction parallel to the metal surface at a distance from it from the above and below.

In Ref [5], in the experimental set-up of matter wave tractor beam the scatterer was modelled by a short-range spherical potential energy. In our work we have used Tang-Toennies potential energy [18] to model a realistic scatterer. Using spherical particles in the simulation, the lateral force was achieved by placing the metal surface vertically on one side of the scatterer. The left-right asymmetry was created by the background potential energy resulting from the interaction between the incident particles and the surface. The demonstration of the experiment, both analytically and by simulation, confirms the presence of reversible lateral force.

Adsorption of particles on solid surfaces can be divided into two categories: physisorption, caused by Van der Waals force, and chemisorption, caused by chemical bonds between adsorbate particles and the adsorbent solid [19]. Since one of our objectives is to ensure that there is no chemical bond between the incident particles and the metal, the focus of this paper was on physisorption. We have chosen both the incident particle and the scatterer to be noble gases further ensuring no chemical reaction in the process.

For our proposed set-up, when the target atom was significantly away from the metal, such that their potential energies had zero overlap, it experienced a lateral force away from the metal in presence of quantum mechanical matter wave. Due to the minimal overlap scenario, the incident beam was first scattered by the target atom. The initial incident beam did not create any lateral movement of the scatterer because of symmetric scattering in both positive and negative x-directions. However, the portion of the scattered wave travelling towards the direction of metal (in the –ve x-direction) got reflected back by the infinite barrier of physisorption potential energy. This reflected wave caused a second scattering with the target atom and thus exerted a pushing force on the target atom in the direction away from the metal.

In contrast, when the target atom was at or within the adsorption length from the metal, their potential energies overlapped significantly. The infinite barrier of the physisorption potential covered a significant portion of the volume of the scatterer potential energy region (c.f. Fig. 1(b)). As a result, there was significant scattering of the incident particles, from the far side of the scatterer in the direction away from the metal (+ve x-direction). On the near side of the metal, less number of incident particles were scattered from the scatterer resulting in a less scattering toward the metal. The total momentum of the scattered wave was, therefore, away from the metal. To conserve momentum, the scatterer gained an equal and opposite momentum towards the metal. However, the scatterer gains energy through the process. If this energy exceeds the adsorption binding energy for adsorption, it should be able to escape the surface. For atomic-sized particles (as incident particles) which share similar adsorption over a metal situation, we adopt a method by utilizing quantum mechanical matter wave (instead of light beams), which may be exploited to extract both chemisorbed and physiosorbed atoms, depending on the strength of the applied force.

With the shrinking of device design rules (Ref [20]), the requirement for wafer cleaning is becoming crucial. The lengthy cleaning process takes up to 30-40% of the fabrication process [20]. So a faster method of cleaning is required. A good aspect of this process is that the incident waves do not need to be localized (only their directions have to be maintained) and can clean a wide surface area. In order to use this phenomenon for cleaning the surface, we can keep the substrate vertically or horizontally and in an upside-down orientation. This assures that the impurities do not fall back onto the surface due to gravitational force, which would be the case for a horizontal upside up surface.

**Set-up and Method**

Consider scattering of incident helium atoms by some heavy atom in the presence of external surface energy. Figure. 1 features the schematic diagram of our proposed model, where the atom is situated at a distance $d$ from a semi-infinite surface's outermost layer. The center of the atom is taken as the origin of the coordinate system. We impinge two plane matter waves of the same energy E on the scatterer from the $z$-direction and the $-z$-direction, respectively. Instead of dealing with the entire wave packets, we are considering a plane wave component of energy E of the wave packet. In this experiment, we have chosen the incident particles to be Helium-4, the scatterer as the atoms of the Xenon gas, and the metal as Cu. Simulation data is generated for $d = 0$ Å to $d = 10.0$ Å. Furthermore, three specific scenarios have been analyzed for $d = 10.0$ Å (an arbitrarily small distance has been taken ensuring minimum overlap between the He-Cu and He-Xe interaction energies for simplified analysis), $d = 3.6$ Å (adsorption bond length for Xe-Cu taken from Ref. [21]), and $d = 1.0$ Å (the choice of distance is based on the physical explanation, which is discussed in details in the next section).

Unlike in Ref. [5], we cannot use the Center of Mass reference frame, as an external force acts on the particles resulting from the space-varying physisorption potential energy. We will, therefore, use Laboratory reference frame. As the target atom vibrates with a small vibration energy [21], it possesses small kinetic energy but no total momentum. If the vibrational motion is ignored, the target particle can be considered motionless. Hence, in the simulation we solved the Schrödinger equation for the incident particles under the influence of stationary scatterer and the metal surface.

The data required to model the set-up are, therefore, the effective mass $m$ of particles in the incident beam, the energy $E$ of the incident beam, and the total potential energy of the system. A constant effective mass of the incident particle is assumed throughout the entire experiment. The beam's energy (25 meV) is given in the same way as in Ref. [5], where a coherent beam with thermal energy at room temperature was provided using an energy selector. The energy is varied from 0 meV to 40 meV for further data analysis. Since the incident particle's energy is a fixed parameter here and the potential energy is time-independent, it was possible to run a stationary study (the supplement includes further information regarding simulation).

The potential energy of the whole set-up has been divided into two parts: the interaction energy experienced by the incident particles due to the surface (adsorption energy, shown in Figure. 3) and that due to the scatterer (He-Xe potential energy). For the second part, Tang-Toennies' potential energy is taken into account (shown in Fig. 2) [18]. The shape of the interaction is considered to be a step approximation of the potential energy curve (this model reduces the complexity of the calculation and allows effective implementation of quantum mechanical (QM) stress tensor [5,22,23] in COMSOL Multiphysics [24]). As the Hamiltonian of a system with time independent potential energy is the sum of the kinetic energy of the incident particles and the total potential energy, we added both the scalar potential energies as functions of position. The net potential energy model of the set-up is shown in Fig. 4.

The stress tensor has been used to calculate the time averaged forces acting on a particle. A similar method is used in optics to calculate the time-averaged optical forces using Minkowski stress tensor [10, 25]. As we are considering matter wave as the incident wave, we are using the QM stress tensor instead of the Minkowski stress tensor. The force density vectors in the $x$, $y$, and $z$-directions were calculated and integrated at $r = 9BR^+$ distance (where $BR^+$ is the Bohr radius) [2] to produce the respective components of the time averaged force acting on the particle (further details are given in the supplement). Since we considered the He-Xe potential energy to be negligible or zero for a distance $r > 9BR$, any closed surface at a distance $r > 9BR^+$ can be used for surface integration.

$$\langle \vec{F}_{Total} \rangle = - \oint \langle \overleftrightarrow{T} \rangle . d\vec{s} \qquad (1)$$

$$\overleftrightarrow{T} = \frac{\hbar^2}{2m} \left( \vec{\nabla} \psi \otimes \vec{\nabla} \psi^* + \vec{\nabla} \psi^* \otimes \vec{\nabla} \psi \right) + \overleftrightarrow{I} \left( (E - U)|\psi|^2 - \frac{\hbar^2}{2m} \left| \vec{\nabla} \psi \right|^2 \right) \qquad (2)$$

where $\psi$ is the total stationary wave function, $m$ is the mass of the incident particles, $E$ is the energy of the incident beam, $U$ is the potential energy, $\overleftrightarrow{I}$ is the unity tensor, and $\langle \rangle$ denotes time average [3]. Using (1) and (2), the following formula [5] is obtained:

$$\vec{F} = - \frac{\hbar^2}{2m} \left[ r^2 \oint \left( 2 \, Re \left[ \vec{\nabla} \psi \left( \vec{\nabla} \psi^* . \hat{n} \right) \right] + \hat{n} \left( k_0^2 |\psi|^2 - \left| \vec{\nabla} \psi \right|^2 \right) \right) d\Omega \right] \qquad (3)$$

where $\hat{n}$ is the outward unit vector. From here, it is evident that $\left| \vec{F} \right| \propto |A|^2$ where $A = $ amplitude of the incident beam.

By applying the QM stress tensor method, the time-averaged pulling force for the set-up of ref [5] has been verified as a first exercise using full-wave simulation method of COMSOL (detailed results have been given in the supplement). After that we adopt our set-up which is quite different from the set-up presented in ref [5]. A short description of our set-up along with the required boundary conditions applied in the COMSOL software (for the full-wave simulation) is given in the supplement of this article.

**Physical Explanation and Analytical Result**

The overall lateral force acting on the scatterer can be explained by three possible scenarios: (A) When there is no overlap between the target atom's potential energy and the infinite barrier of the physisorption potential energy; (B) When there is an overlap, but the direction of a significant portion of the incident beam (before interaction with the target atom) remains unaltered by the influence of the physisorption potential energy; and (C) when there is a greater overlap so the direction of a significant portion of the incident beam (before interaction with the target atom) changes by the influence of the physisorption potential energy. Depending on the amount of contribution of each scenario, the lateral force's direction and magnitude change.

The first scenario can be divided into three consecutive events: (i) the single-channel scattering of incident particles by the scatterer (since the incident particles are far away from the metal-incident particle potential energy barrier), (ii) subsequent interaction of the waves scattered in the previous event, with the physisorption potential energy of the incident atom-metal, and (iii) finally, a second single-channel scattering of the resulting reflected wave by the metal with the scatterer Xe atom. Although time-independent scattering theory can be applied in the case of our time-independent potential energy, the limitations of the available analytical method for such asymmetric potential energy lead us to divide the events sequentially.

For a simplified calculational procedure, we are considering the value of $d$ equal to 10 Å, such that there is no overlap between the physisorption potential energy of the metal-incident atom and the scatterer-incident atom potential energy. The first event is a 3D scattering of plane waves by a spherical potential energy, allowing us to use the Partial Wave Analysis method for deriving the total wave function. The scattering amplitude is then used to find the portion of the wave scattered toward the metal. Our target is to find non-zero scattering amplitude for $\theta = \pi/2$, where $\theta$ is the polar angle with respect to the incident beam coming from the $-z$ direction in the $x$-$z$ plane.

The incoming plane waves can be written in terms of spherical waves using Rayleigh's formula as

$$\psi_{inc}(r,\theta) = A\,e^{ikz} + A\,e^{-ikz} = \sum_{l=0}^{\infty} A i^{l}(2l+1)\,j_{l}(kr)\,P_{l}(\cos\theta) + \sum_{l=0}^{\infty} A i^{l}(2l+1)\,j_{l}(-kr)\,P_{l}(\cos(\pi-\theta)) \quad (4)$$

where $A$ is the amplitude of the incident wave, $l$ is the orbital angular momentum quantum number, $j_l$ is $l$-th order spherical Bessel function, and $P_l$ is $l$-th order Legendre polynomial. Since our incident wave function is independent of $\phi$ ($z = r\cos\theta$), only $m = 0$ term survives; therefore, we have omitted all the $m \neq 0$ terms.

Expanding Eq. (4) in terms of Hankel's function of the first and second kinds, which go to $e^{+ikr}/kr$ and $e^{-ikr}/kr$, respectively for $kr \gg 1$, and including phase shift of the outgoing waves due to the spherical potential energy, we get:

$$\psi(r,\theta) = A\left[e^{ikz} + e^{-ikz} + \left(\sum_{l=0}^{\infty}(2l+1)\frac{1}{k}e^{2i\delta_l}\sin(\delta_l)P_l(\cos\theta) + \sum_{l=0}^{\infty}(2l+1)\frac{1}{k}e^{2i\delta_l}\sin\delta_l P_l(\cos(\pi-\theta))\right)\frac{e^{ikr}}{r}\right] \quad (5)$$

From Eq. (5), the overall scattering amplitude for the first scattering event can be written as (the derivation is included in the supplement)

$$f_{overall}(\theta) = f(\theta) + f(\pi-\theta) = \sum_{l=0}^{\infty}(2l+1)\frac{1}{k}e^{2i\delta_l}\sin(\delta_l)\left[P_l(\cos\theta) + P_l(\cos(\pi-\theta))\right] \quad (6)$$

For this, the $l$-th partial wave amplitude relation with $\delta_l$ is given by the following relation:

$$a_l = \frac{1}{2ik}\left(e^{2i\delta_l} - 1\right) = \frac{1}{k}e^{2i\delta_l}\sin\delta_l \quad (7)$$

To get the scattering amplitude, we will apply the required boundary conditions. The scatterer potential energy is modeled as follows:

$$V(r) = \begin{cases} \infty & (r < b) \\ -V0 & (b \leq r \leq a) \\ 0 & (r > a) \end{cases}$$

where $V_0 = 2.458$ meV, a=9×Bohr Radius, and b= 6.697×Bohr Radius for our model. Applying the following three appropriate boundary conditions at $r = a$ and $r = b$ gives Eq. (8) and (9) (detailed calculation is given in the supplement).

$$\psi(r = a^-) = \psi(r = a^+) \quad \text{(i)}$$

$$\frac{\partial \psi}{\partial r}\bigg|_{r=a^-} = \frac{\partial \psi}{\partial r}\bigg|_{r=a^+} \quad \text{(ii)}$$

$$\psi(r=b) = 0 \quad \text{(iii)}$$

$$P/Q = R/S \quad (8)$$

where,

$$P = i^l j_l(ka) + k i^{l+1} a_l h_l^{(1)}(ka) + i^l j_l(-ka)(-1)^l + k(-i)^{l+1} a_l h_l^{(2)}(-ka)$$

$$R = i^l j_l(k'a) + k' i^{l+1} a'_l h_l^{(1)}(k'a) + i^l j_l(-k'a)(-1)^l + k'(-i)^{l+1} a'_l h_l^{(2)}(-k'a)$$

$$Q = \frac{1}{2}k[j_{l-1}(ka) - j_{l+1}(ka)]i^l + k i^{l+1} a_l \frac{1}{2} k\left[h_{l-1}^{(1)}(ka) - h_{l+1}^{(1)}(ka)\right]$$
$$- i^l \frac{1}{2}k[j_{l-1}(ka) - j_{l+1}(ka)](-1)^l - k(-i)^{l+1} a_l \frac{1}{2}k\left[h_{l-1}^{(2)}(-ka) - h_{l+1}^{(2)}(-ka)\right]$$

$$S = i^l \frac{1}{2}k'\left[j_{l-1}(k'a) - j_{l+1}(k'a)\right] + k' i^{l+1} a'_l \frac{1}{2}k'\left[h_{l-1}^{(1)}(k'a) - h_{l+1}^{(1)}(k'a)\right]$$
$$- i^l \frac{1}{2}k'\left[j_{l-1}(k'a) - j_{l+1}(k'a)\right](-1)^l - k'(-i)^{l+1} a'_l \frac{1}{2}k'\left[h_{l-1}^{(2)}(-k'a) - h_{l+1}^{(2)}(-k'a)\right]$$

$$a_l = \frac{-\left[j_l(kb) + j_l(-kb)(-1)^l\right]}{ik'\left[h_l^{(1)}(k'b) - h_l^{(2)}(-k'b)\right]} \quad (9)$$

where k'=$\sqrt{2m(E-(-V_0))}/\hbar$. We used Eq. (8), and (9) to get $a_l$, finally $f_{overall}(\theta = \pi/2)$ using Wolfram Mathematica. The $l$-th terms are taken up to $ka = 33$ (where $k = \sqrt{2mE}/\hbar$ and $E = 25$ meV; the calculation is shown in the supplement). This is because it is a high-energy scattering for which $l \leq kr$ terms should be taken into account (where $r$ is the radius of the spherical potential energy).

The scattering amplitude for $\theta=\pi/2$ is therefore:

$$f(\theta = \frac{\pi}{2}) = \sum_{l=0}^{33} [(2l+1)e^{2i\delta_l}\sin(\delta_l)P_l(0)[\frac{1}{k} + \frac{(-1)^l}{k}]] \quad (10)$$

Using Eq. (10), we found the scattering amplitude at $\theta = \pi/2$ for step approximated He-Xe potential energy model and $E = 25$ meV to be 0.824716. Similarly, calculated scattering amplitude is non-zero for other incident beam energies for which force is in the $x$ direction (Mathematica results for different energies have been provided in the supplemental material). Calculating the scattering amplitude is necessary as we have seen in Eq. (3) that the scattering force is proportional to the square of the amplitude of the incident beam.

The shape of the interaction energy graph for physisorption is an essential part of the second event and can be explained by Van der Waals attraction and Pauli repulsion. As the electron wave functions of the adsorbent and adsorber overlap, the attractive Van der Waals force creates potential wells above the surface of depth of around a few meV. On the other hand, the system's energy increases due to the orthogonality of the wave functions according to the Pauli exclusion principle, which states that no two fermions (in this case, the electron clouds of the particles and the metal atoms) can be in the same quantum state. Pauli exclusion and consequent repulsion are strong for noble gases, dominating the short-range interaction between our incident particles and the metal surface. This combination of short-range Pauli repulsion and long-range Van der Waals attraction results in a one-side steep or bounded and one-side open or flat curve of potential energy which allows the scattered wave toward the metal to be reflected back from the bound side and escape through the open side.

For simplicity, only the portion of the scattered plane waves (from the previous event) with k vector in $-x$ direction is considered, taking into account their maximum contribution to the current event.

$$\psi_{scattered} = \frac{f\left(\theta = \frac{\pi}{2}\right)}{|x|} e^{-ikx}$$

The reflected wave will be in the form:

$$\psi_{reflected} = Be^{ikx} = R \frac{f\left(\theta = \frac{\pi}{2}\right)}{|x-\varepsilon|} e^{i\delta} e^{ikx} \qquad (11)$$

where $\varepsilon = 10^{-10}$ is a small distance (used in order to avoid numerical error), $R$ is the reflection coefficient for the reflection of He from the He-Cu potential energy barrier, and $\delta_l$ is the phase shift of the reflected wave due to this potential energy. Using 1D simulation in COMSOL, the numerical value of $R$ is computed to be 0.99813.

$\psi_{reflected}$ is the incident beam for the second scattering event. When incident plane waves impinge on the scatterer from a single direction, only pushing force in the direction of incidence is exerted on the scatterer [5]. Similarly, for this case, the $x$-component of the momentum of the scattered wave from the second scattering event will be less than that of the incident beam ($\psi_{reflected}$), and hence due to momentum conservation, the scatterer gains momentum in the positive $x$-direction. Therefore, the lateral force on the scatterer will be in the $x$ direction. Using the value of R and $f\left(\theta = \frac{\pi}{2}\right)$, the magnitude of the lateral force for E = 25 meV should be $\left(R \times \frac{f\left(\theta=\frac{\pi}{2}\right)}{|x-\varepsilon|}\right)^2 \times Fz \approx$ $0.68 Fz$, where Fz is the pushing force exerted by a single plane matter-wave (cf. Table 1).

For the second scenario, since the potential energy modeling is NOT spherically symmetric, the analytical solution of time-independent Schrödinger equation (TISE) is intractable. However, the direction of the net force on the scatterer can be obtained due to significant rightward scattering (in the x-direction) compared to insignificant leftward scattering (as the potential barrier covers the majority of the left side volume of the scatterer's potential energy). Since the final total momentum of the scattered wave is in the x-direction, the scatterer gains equal and opposite momentum to conserve total zero lateral momentum of the system. The lateral force on the scatterer will, therefore, be in the $-x$ direction.

For the third scenario, the analytical solution of TISE is also intractable. In this case, a significant portion of the incident beam goes through a change of direction before getting scattered by the target atom. As the $x$-axis gradient of He-Cu physisorption potential energy is higher on the left side of the minimum He-Cu potential energy than to the right, the incident beam changes direction more to the right due to gradient force ( $\vec{F} = -\vec{\nabla} U$ ). Therefore, it exerts a lateral force to the $x$ direction.

**Results and Discussion**

We verified that for symmetric set-up (in the absence of the Copper surface), there is no lateral force. The set-up is shown in Fig. 5, and the results are shown in Table 1 for $E = 25$ meV.

| Incident Beam Direction | Fx (N) | Fy (N) | Fz (N) |
|---|---|---|---|
| $-z$ and $z$ | 2.07× $10^{-22}$ ≈ 0 | -1.09× $10^{-22}$ ≈ 0 | -1.22× $10^{-23}$ ≈ 0 |
| $z$ | 3.11× $10^{-24}$ ≈ 0 | -3.64× $10^{-23}$ ≈ 0 | 1.79× $10^{-19}$ |

Table 1: This table represents the force components acting on the scatterer in the absence of the metal surface for $E = 25$ meV. From here, we can conclude that $F_x$, $F_y$, and $F_z$ can be considered zero for the incidence of plane waves from the $z$ and $-z$ directions because these values are insignificant compared to $F_z$ for the set-up where the incidence of plane waves is from the $z$ direction (any value in the scale of $10^{-21}$ or less is considered negligible or zero). The non-zero values of $\langle F_x \rangle$, $\langle F_y \rangle$ and $\langle F_z \rangle$ for the two-beam scenario are due to small computational error.

The set-up for simulation in the presence of a Copper surface and single incident beam is given in Fig. 6, and the results are given in Fig. 7(a) for $E = $ 25 meV. From the figure, it can be observed that the scatterer experiences a significant lateral force in the presence of the metal surface and a pushing force is also present.

Since we are looking for a lateral force, we introduce a second beam incident in the $-z$ direction resulting in a pair of counter-propagating incident beams. The results for simulation in the presence of Copper surface and counter-propagating incident beams are given in Fig. 7(b) for $E = $ 25 meV. There is no pushing force as seen in the figure. In the previous section, we saw that the magnitude of lateral force for $d = 10.0$ Å should be $0.68 Fz$. However, due to the assumptions made regarding the reflection wave, this magnitude is higher than the lateral force obtained from the simulation. The force variation due to change in d for different incident energies E is illustrated in Fig. 8. In this figure, the pattern of the direction of lateral force as d increases can be observed to be from x direction to -x direction to x direction for certain energy values. For higher energies, the lateral force goes from -x direction to x direction with increase in d. Point A (E=25 meV and d=10Å) represents the force from the first scenario, therefore, the lateral force is in the x direction. Similarly, Point B represents the force where the second scenario dominates, therefore, the lateral force is in the -x direction. As discussed in the third scenario, it can be seen that at point C (E= 10 meV and d=1.0Å), the lateral force is away from the metal. Table 2 represents the simulation result for force components for Xe-Cu adsorption bond length distance.

| Distance from the metal surface | Fx (N) | Fy (N) | Fz (N) |
|---|---|---|---|
| d=3.6 Å | $-1 \times 10^{-19}$ | $-1 \times 10^{-22} \approx 0$ | $-1 \times 10^{-21} \approx 0$ |

Table 2: This table represents the force components for scatterer distance of 3.6 Å from the metal and the incidence of plane waves from both the $z$ and $-z$ directions. From here, we can conclude that a lateral force in the $-x$ direction is experienced by the scatterer.

For the adsorption case, the lateral force acts towards the surface (in $-x$ direction). As far as our concern is, this force causes the scatterer wave packet to move toward the surface by gaining a wave vector component in the $-x$ direction. However, due to this momentum transfer, the kinetic energy rises, and therefore the thermal energy of the scatterer rises (initially the scatterer's energy is negative of binding energy due to the exothermic process of adsorption [27]). If the scatterer's gain in thermal energy exceeds the adsorption binding energy, it will be desorbed. Furthermore, as the variation of the Xe-Cu potential energy (see Ref. [21]) with respect to the distance from Cu in the x direction is similar to that of He-Cu (shown in Fig. 3), the scatterer will be able to escape through the open side. Therefore, regardless of the x-axis direction of lateral force, the scatterer will completely leave the surface if its energy gain becomes greater than the binding energy. Moreover, to further ensure cleaning, vacuum pumping can be introduced [2].

The initial total momentum of the incident beams is zero as the beams are counter-propagating. The transferred momentum from the incident particle to the scatterer would be maximum if the model is such that the scattering amplitudes for either $\theta = \pi/2$ or $\theta = -\pi/2$ are 1 (i.e., the maximum value, as the scattering amplitude is a probability amplitude) for both the incident beams. The total transferred momentum $q$ can, therefore, vary from 0 to $q = 2\hbar k$ where 2 is the total scattering amplitude contributing to the lateral force, and $\hbar k$ is the momentum of each incident beam. Using the following relation between incident energy and scatterer energy (using $E = p^2/2m$), we can find the maximum transferred energy to the scatterer as:

$$E_{max} = \frac{4m_{He}}{m_{Xe}} E = \left(\frac{16}{131} \times 25\right) meV \tag{12}$$

From the above relation, we get the maximum energy transferred for our set-up is around 3 meV. However, if the incident particle energy is high (for example, around 2 eV), $E_{max}$ would be around 244 meV, which exceeds the adsorption binding energy of Xe-Cu (111) (173-200 meV) [21]. Hence, depending on the incident particle energy and the scattering amplitude, it is possible to overcome the adsorption binding energy. Moreover, no Helium would be adsorbed as its binding energy is very small (4 meV) compared to its incident energy (25 meV). The shape of the incident particle-metal physisorption energy makes such an incident particle a more suitable candidate for desorption than electrons. This is because the surface electron density does not change in the case of ESD [28]. The incident particles are unable to tunnel to the surface.

A real-world example where this technique can be utilized is to clean graphene surfaces. The interaction of some atomic-sized particles with the graphene surface also has similar adsorption energy curves as He-Cu [29]. Hence, such atomic-sized particles can be used as incident particles to remove unwanted particles from graphene before using it as a semiconductor for modern electronics [30].

**Conclusion**

In conclusion, we have presented a simple configuration in which atomic-sized particles can experience reversible lateral force in presence of quantum mechanical matter wave resulting from the presence of surface energy. The configuration has been set to reflect a real-world situation and has been verified using full-wave simulation. When the potential energies of the target atom and the surface had minimal overlap, the direction of lateral force was away from the metal. Due to minimal overlap, the target atom first scattered the incident beam. The scattered wave got reflected back from the infinite barrier of physisorption potential energy and triggered a second scattering with the target atom. On the other hand, when the target atom was closer to the metal such that the infinite barrier of the physisorption potential covered the majority of half of the volume of the scatterer potential energy region, there was more scattering away from the metal than toward it. Therefore, the scatterer experienced a force toward the metal. Moreover, when the infinite barrier covered the majority of the volume of the scatterer potential energy region (when the target atom is even closer to the surface), the direction of a significant amount of incident beam changes to the direction away from the metal due to the gradient force by physisorption potential energy. This causes the incident beam to exert a pushing force on the scatterer away from the metal. This manipulation can be utilized in fields where any kind of background potential energy is involved. Additionally, desorption without affecting the substrate can be possible by using an incident particle that shares similar potential energy with the surface as He-Cu.

**Acknowledgement**
M.R.C. Mahdy acknowledges the support of NSU CTRGC grant 2021-22 and 2022-23 of North South University.

**Supplementary information**

**Figures and Captions List**

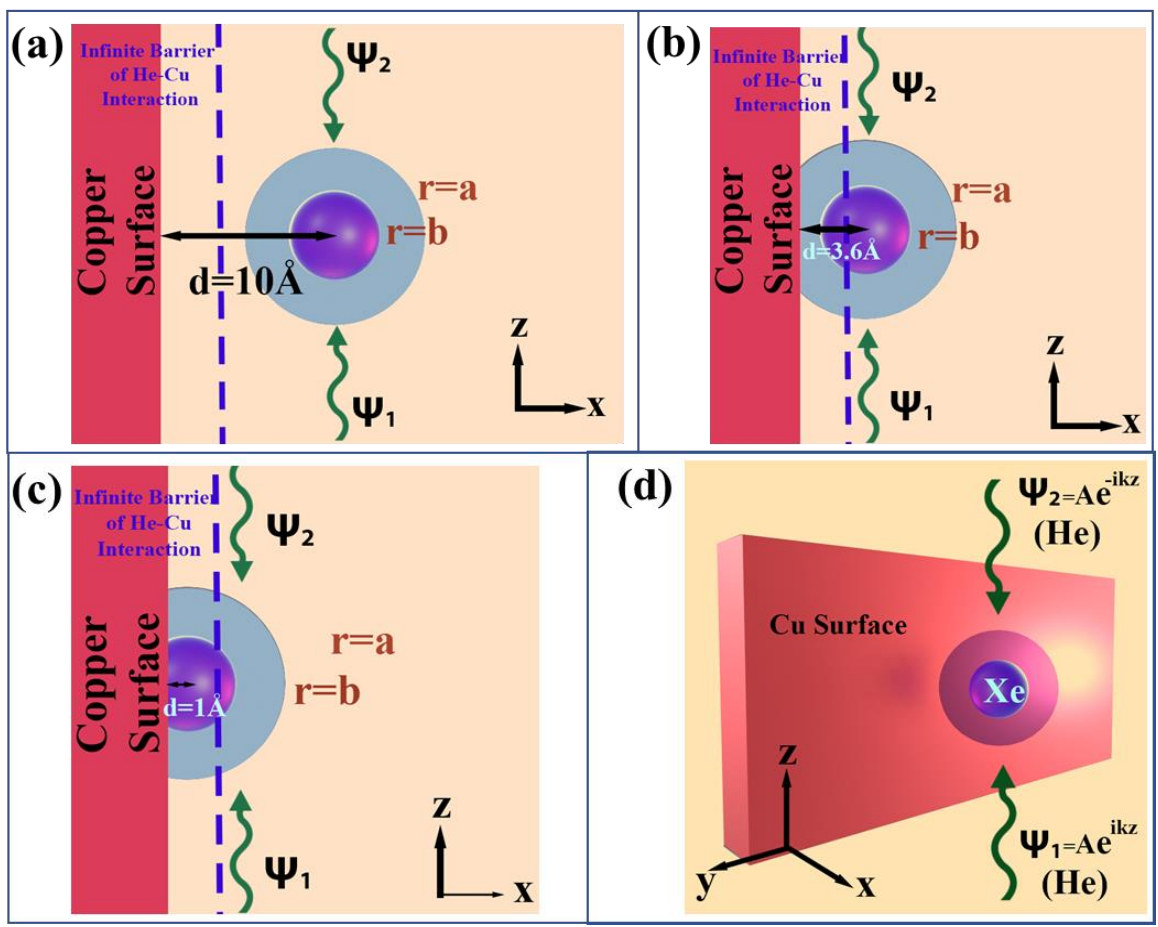

Fig. 1 (a), (b), and (c) illustrate the $x - z$ plane view of the experimental set-up for $d = 10.0$ Å, $d = 3.6$ Å, and $d = 1.0$ Å, respectively, where $d$ is the distance between the scatterer (Xe) and the metal (Cu) surface. Two plane matter waves (He) impinge on the scatterer in the $z$ and $-z$ directions. The He-Cu physisorption potential energy is infinite on the right-side region of the dashed line. However, the total variation in physisorption potential energy is not shown here. The potential energy for He-Xe is infinite for $r < b$ (dark blue region), $-2.458$ meV for $b \leq r \leq a$ (semi transparent region), and 0 for $r > a$, where $b = 6.697 BR$ and $a = 9 BR$, where $BR =$ Bohr radius. b) The figure shows the geometry of the model in 3D.

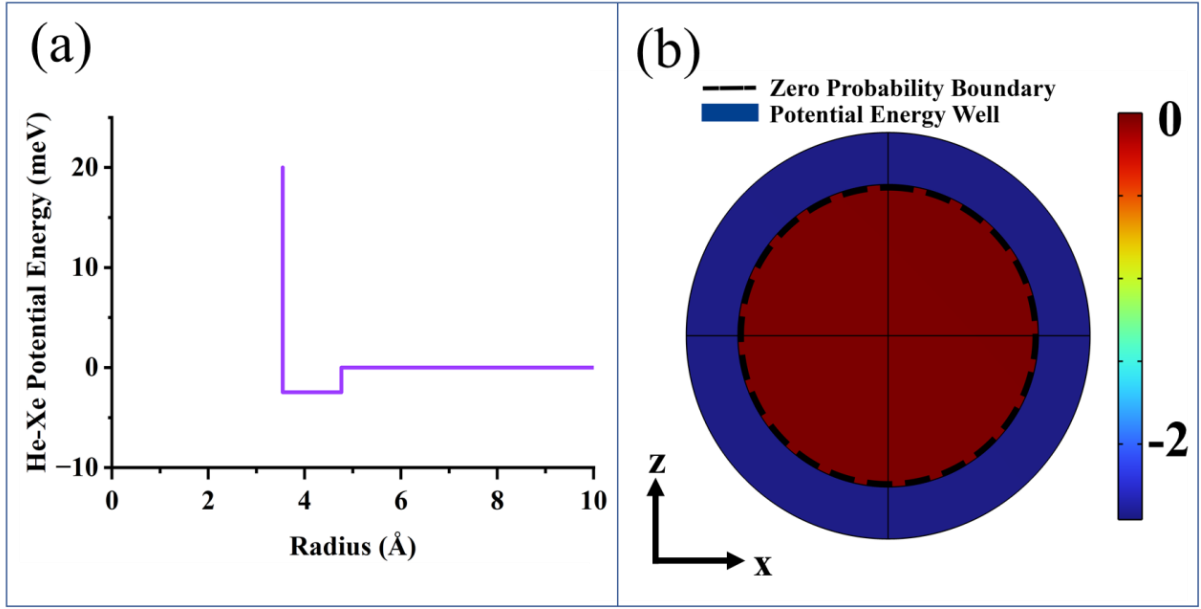

Fig.2 a) The graph represents the step approximation of the potential energy between He and Xe atoms collected from [14]. b) The figure shows the potential energy modeling between He-Xe in COMSOL. The potential energy is infinite inside the circular Zero Probability Boundary.

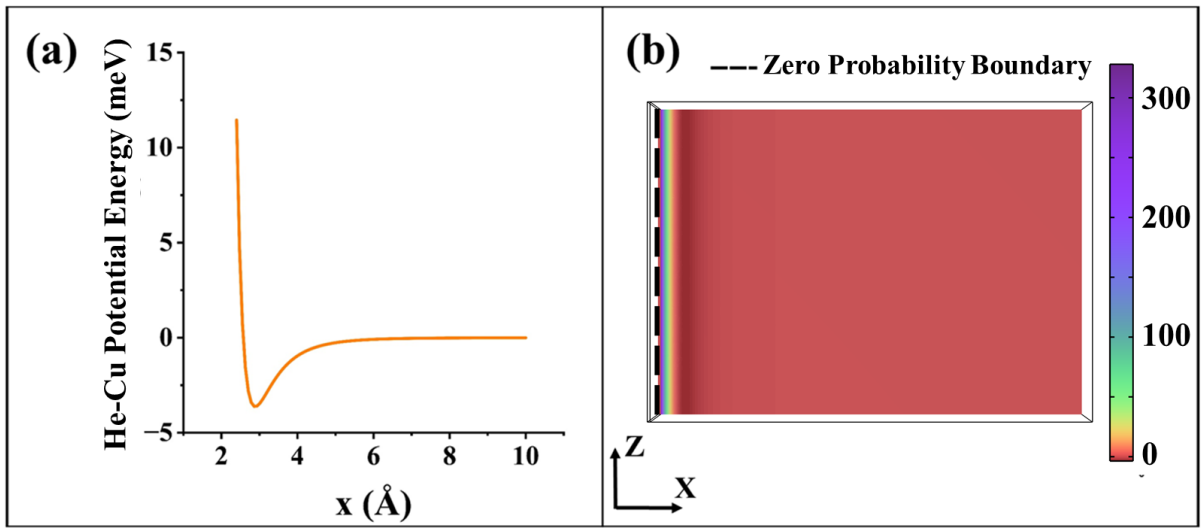

Fig.3 (a) This graph represents the experimental potential energy of our incident beam particle (He) in relation to its separation from the Cu surface [26]. Using the information from (a), the potential energy of the incident beam under the influence of the metal surface is modeled. (b) This figure shows the modeling in COMSOL. The potential energy is infinite on the left region of the straight line representing Zero Probability Boundary which is at a distance of 2.0 Å from the Cu surface.

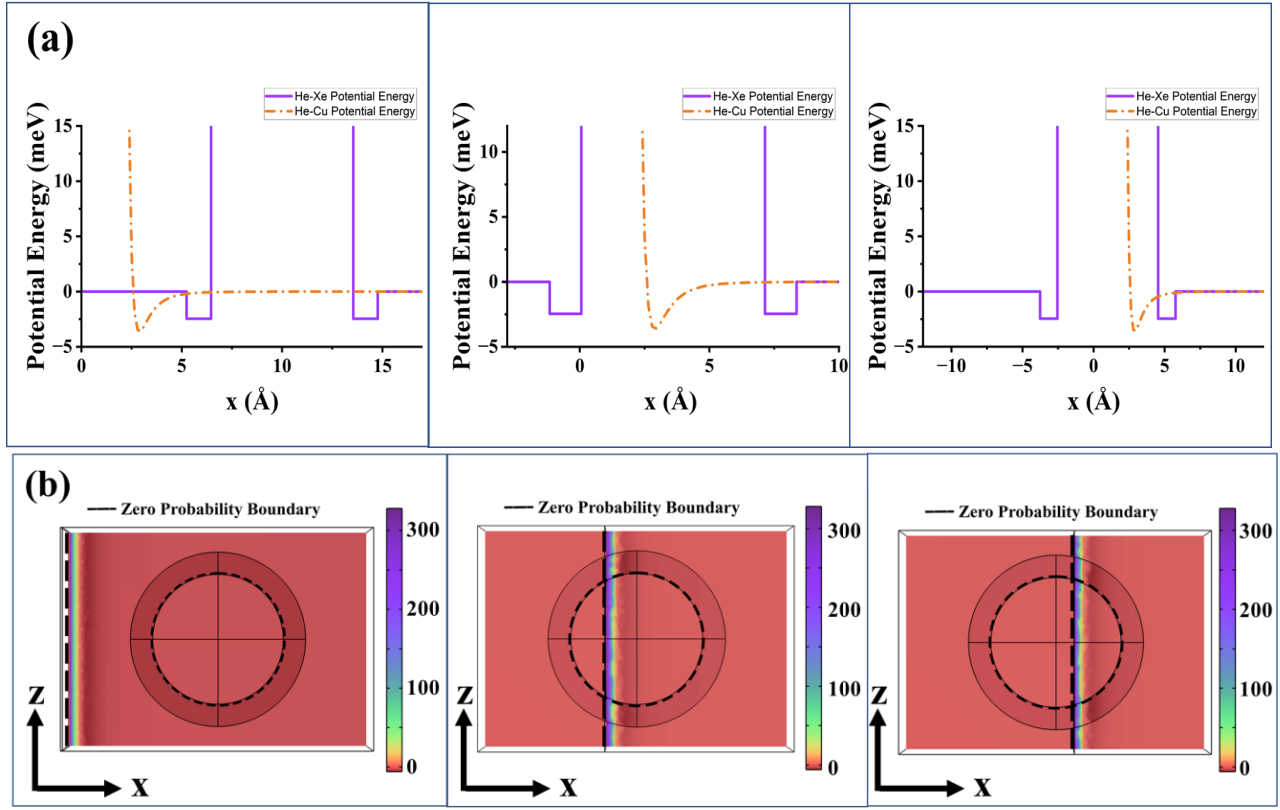

Fig.4 a) The graphs illustrate the potential energy overlap shown in Fig 2-3 for $d = 10.0$ Å, $d = 3.6$Å, and $d = 1.0$ Å, respectively. b) The figures show the $x - z$ view of the net potential energy modeling in COMSOL for $d = 10.0$ Å, $d = 3.6$ Å, and $d = 1.0$ Å, respectively. The potential energy is infinite inside the circular Zero Probability Boundary and on the left region of the straight line of Zero Probability Boundary.

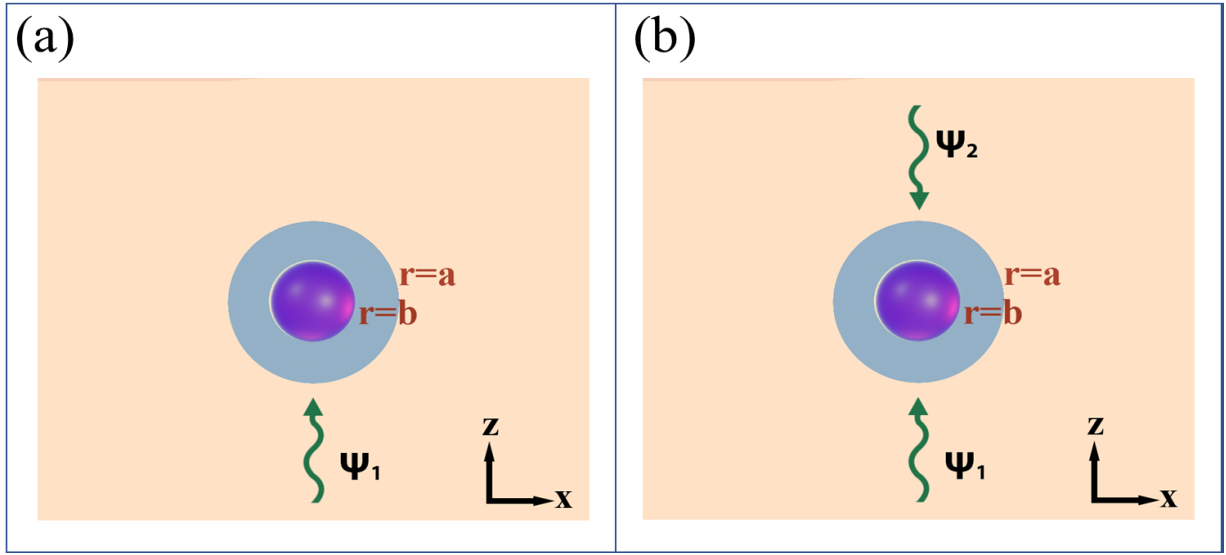

Fig.5 Figures (a) and (b) show the set-up for the counter-propagating incident beam and single incident beam, respectively, in the absence of the Copper surface. In both cases, there was no lateral force experienced by the scatterer.

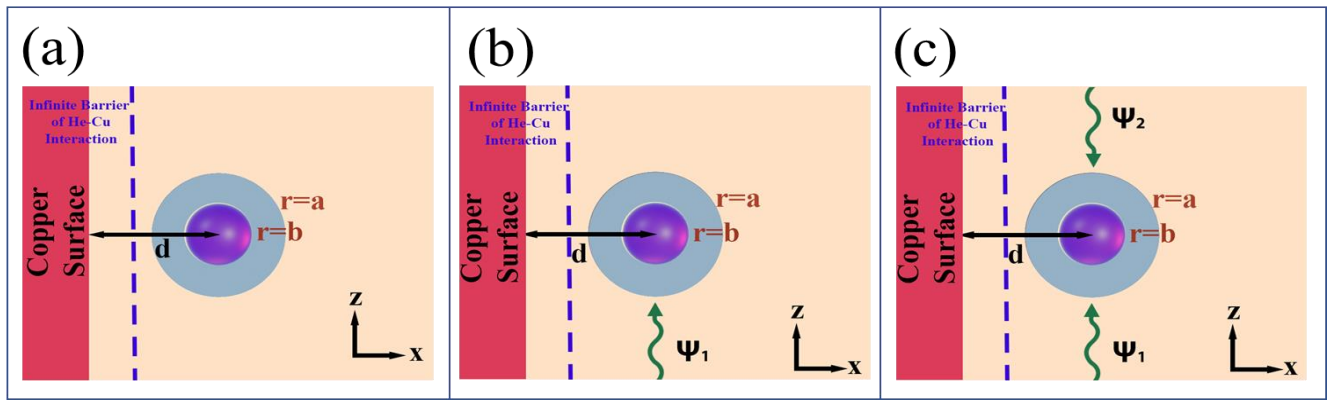

Fig. 6 Figures (a), (b), and (c) illustrate the $x - z$ plane view of the experimental setup for no incident beam, single incident beam, and counter-propagating incident beams, respectively, where $d$ is the distance between the scatterer (Xe) and the metal (Cu) surface. (a) No force was observed for the no incident beam scenario b) Lateral force, as well as pushing force in the $z$ direction was observed for a single incident beam. (c) Only lateral force was observed in this case, as the vertical forces got canceled out. To ensure only lateral force, we chose this as our main set-up.

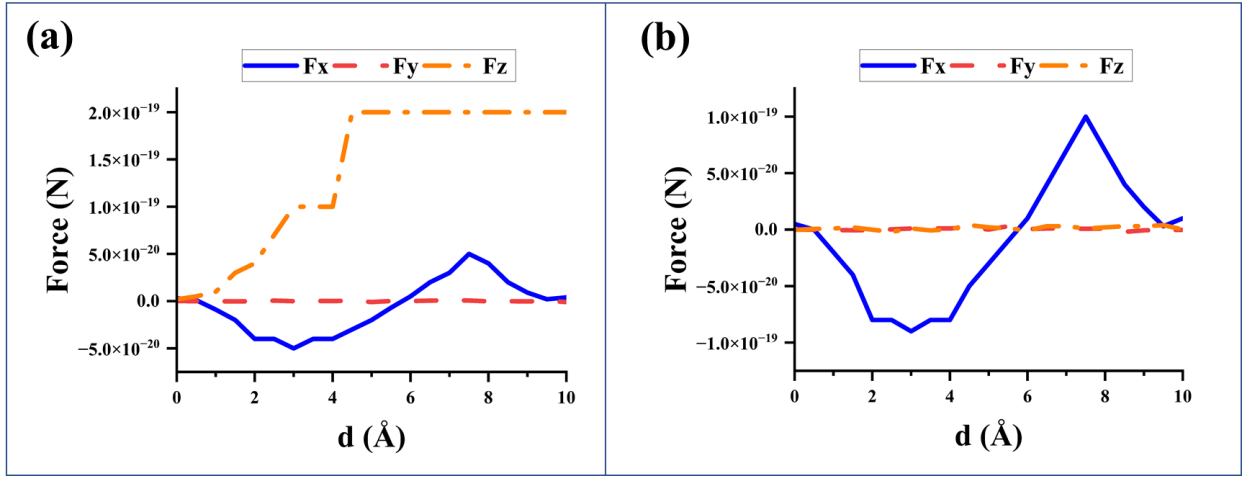

Fig. 7 a) This graph represents the force components on the scatterer for variation in distance from the scatterer to the metal (i.e. $d$) and the incidence of plane matter waves from the $z$ direction only (from a single direction). From here, we can conclude that there is a net lateral force on the scatterer in the $x$ direction for small $d$ and in the $-x$ direction for large $d$. Moreover, a pushing force in the $z$ direction is also observed. b) This graph represents the force components on the scatterer for variation in $d$ and the incidence of plane matter waves from $z$ and -$z$ directions (counter propagating plane matter waves). From here, we can conclude that there is a net lateral force (the most dominant component of force among all the components) acting on the scatterer in the $x$ direction for small d and in the $-x$ direction for large $d$.

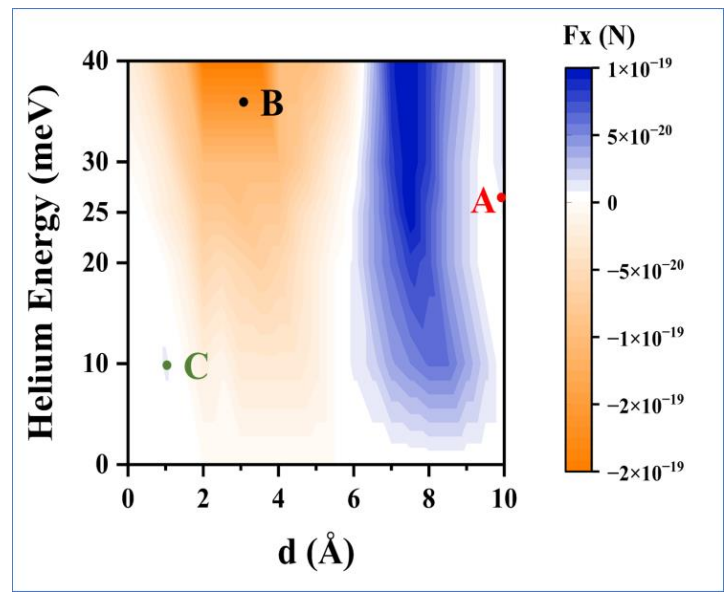

Fig. 8 The graph represents the variation in $F_x$ (lateral force) for different values of $d$ (distance between the center of the target atom and the surface) and $E$ (energy of the incident particles). Points A, B, and C represent the setups where the first, second, and third scenarios dominate, respectively.

# References

[1] Ashkin, Arthur. "Acceleration and trapping of particles by radiation pressure." Physical review letters 24, no. 4 (1970): 156.

[2] Ashkin, Arthur, James M. Dziedzic, John E. Bjorkholm, and Steven Chu. "Observation of a single-beam gradient force optical trap for dielectric particles." Optics letters 11, no. 5 (1986): 288-290.

[3] Ashkin, Arthur, and James M. Dziedzic. "Optical trapping and manipulation of viruses and bacteria." Science 235, no. 4795 (1987): 1517-1520.

[4] Phillips, William D. "Nobel Lecture: Laser cooling and trapping of neutral atoms." Reviews of Modern Physics 70, no. 3 (1998): 721.

[5] Gorlach, Alexey A., Maxim A. Gorlach, Andrei V. Lavrinenko, and Andrey Novitsky. "Matter-wave tractor beams." Physical Review Letters 118, no. 18 (2017): 180401.

[6] Marston PL. Axial radiation force of a Bessel beam on a sphere and direction reversal of the force. J Acoust Soc Am 2006; 120: 3518–3524.

[7] Chen, Jun, Jack Ng, Zhifang Lin, and Che Ting Chan. "Optical pulling force." Nature photonics 5, no. 9 (2011): 531-534.

[8] Novitsky, Andrey, Cheng-Wei Qiu, and Haifeng Wang. "Single gradientless light beam drags particles as tractor beams." Physical review letters 107, no. 20 (2011): 203601.

[9] Brzobohatý, O., V. Karásek, M. Šiler, L. Chvátal, T. Čižmár, and P. Zemánek. "Experimental demonstration of optical transport, sorting and self-arrangement using a 'tractor beam'." Nature Photonics 7, no. 2 (2013): 123-127.

[10] Qiu, Cheng-Wei, Weiqiang Ding, M. R. C. Mahdy, Dongliang Gao, Tianhang Zhang, Fook Chiong Cheong, Aristide Dogariu, Zheng Wang, and Chwee Teck Lim. "Photon momentum transfer in inhomogeneous dielectric mixtures and induced tractor beams." Light: Science & Applications 4, no. 4 (2015): e278-e278.

[11] Sadgrove, Mark, Sandro Wimberger, and Síle Nic Chormaic. "Quantum coherent tractor beam effect for atoms trapped near a nano waveguide." *Scientific reports* 6, no. 1 (2016): 28905.

[12] Krasnov, I. V. "Bichromatic optical tractor beam for resonant atoms." *Physics Letters A* 376, no. 42-43 (2012): 2743-2749.

[13] Sukhov, Sergey, Veerachart Kajorndejnukul, Roxana Rezvani Naraghi, and Aristide Dogariu. "Dynamic consequences of optical spin–orbit interaction." Nature Photonics 9, no. 12 (2015): 809-812.

[14] Rodríguez-Fortuño, Francisco J., Nader Engheta, Alejandro Martínez, and Anatoly V. Zayats. "Lateral forces on circularly polarizable particles near a surface." Nature communications 6, no. 1 (2015): 8799.

[15] Zhang, Tianhang, Mahdy Rahman Chowdhury Mahdy, Yongmin Liu, Jing Hua Teng, Chwee Teck Lim, Zheng Wang, and Cheng-Wei Qiu. "All-optical chirality-sensitive sorting via reversible lateral forces in interference fields." ACS nano 11, no. 4 (2017): 4292-4300.

[16] Wang, S. B., and Che Ting Chan. "Lateral optical force on chiral particles near a surface." Nature communications 5, no. 1 (2014): 3307.

[17] Hayat, Amaury, JP Balthasar Mueller, and Federico Capasso. "Lateral chirality-sorting optical forces." Proceedings of the National Academy of Sciences 112, no. 43 (2015): 13190-13194.

[18] Tang, Kwong Tin and Jan Peter Toennies. "The van der Waals potentials between all the rare gas atoms from He to Rn." Journal of Chemical Physics 118 (2003): 4976-4983.

[19] Kelsall, G. H. "N. SATO Electrochemistry at Metal and semiconductor electrodes." *JOURNAL OF APPLIED ELECTROCHEMISTRY* 30, no. 4 (2000): 515-516.

[20] Wafer Surface Cleaning. "Wafer Surface Cleaning," link: https://www.mks.com/n/wafer-surface-cleaning.

[21] Ruiz, Victor G., Wei Liu, and Alexandre Tkatchenko. "Density-functional Theory with Screened Van Der Waals Interactions Applied to Atomic and Molecular Adsorbates on Close-packed and Non-close-packed Surfaces." *PHYSICAL REVIEW B*, (2016).

[22] Nielsen, O. H., and Richard M. Martin. "Quantum-mechanical theory of stress and force." *Physical Review B* 32, no. 6 (1985): 3780.

[23] Deb, B. M. "On some local force densities and stress tensors in molecular quantum mechanics." *Journal of Physics B: Atomic and Molecular Physics* 12, no. 23 (1979): 3857.

[24] COMSOL Multiphysics Reference Manual, version 6, COMSOL, Inc, www.comsol.com.

[25] Zhu, Tongtong, M. R. C. Mahdy, Yongyin Cao, Haiyi Lv, Fangkui Sun, Zehui Jiang, and Weiqiang Ding. "Optical pulling using evanescent mode in sub-wavelength channels." *Optics express* 24, no. 16 (2016): 18436-18444.

[26] Vidali, Gianfranco, G. Ihm, Hye-Young Kim, and Milton W. Cole. "Potentials of physical adsorption." *Surface Science Reports* 12, no. 4 (1991): 135-181.

[27] Libretexts. "31.6: Atoms and Molecules Can Physisorb or Chemisorb to a Surface." Chemistry LibreTexts. Libretexts, April 1, 2023.
Link:
https://chem.libretexts.org/Bookshelves/Physical_and_Theoretical_Chemistry_Textbook_Maps/Physical_Chemistry_%28LibreTexts%29/31%3A_Solids_and_Surface_Chemistry/31.06%3A_A_Gas_Molecule_can_Physisorb_or_Chemisorb_to_a_Solid_Surface.

[28] Ramsier, R.D., and J.T. Yates. "Electron-Stimulated Desorption: Principles and Applications." Surface Science Reports 12, no. 6-8 (1991): 246–378.

[29] Fenta, A S, C O Amorim, J N Gonçalves, N Fortunato, M B Barbosa, J P Araujo, M Houssa, et al. "Hg Adatoms on Graphene: A First-Principles Study." Journal of Physics: 4, no. 1 (2020): 015002.

[30] "Graphene Semiconductors: Introduction and Market Status." Link: https://www.graphene-info.com/graphene-semiconductor